# Revisiting the Mechanisms of Thermal Transport in Vacancy-Defective Silicon

Xueyan Zhu [1,2,*], Jin Yang [3], J. Shiomi [4], Cheng Shao [5,*]

[1] CAEP Software Center for High Performance Numerical Simulation, Beijing 100088, P.R. China

[2] Institute of Applied Physics and Computational Mathematics, Beijing 100088, P.R. China

[3] ZJU-UIUC Institute, College of Energy Engineering, Zhejiang University, Jiaxing, Haining, Zhejiang Province 314400, P.R. China

[4] Department of Mechanical Engineering, The University of Tokyo, 7-3-1 Hongo, Bunkyo, Tokyo, 113-8656, Japan

[5] Thermal Science Research Center, Shandong Institute of Advanced Technology, Jinan, Shandong Province 250103, P.R. China

---

[*]Corresponding Emails: zhu_xueyan@iapcm.ac.cn, cheng.shao@iat.cn

**Abstract**

Understanding heat conduction in defective silicon is crucial for electronics and thermoelectrics. Conventional understanding relies on phonon gas picture, treating defects as scattering centers that reduce phonon lifetimes without altering frequencies and group velocities. We go beyond phonon gas picture by employing Wigner transport equation to investigate heat conduction in vacancy-defected silicon. Our findings reveal that while thermal conduction in pristine silicon stems mainly from particle-like propagation of vibrational modes, wave-like tunnelling becomes increasingly significant in the presence of vacancies. Contrary to the conventional belief that defects only perturb mode lifetimes, we demonstrate that vacancies also diminish velocity operators — a dominant factor in thermal conductivity reduction, surpassing even the effect of lifetime shortening. Furthermore, incorporating anharmonic frequencies and interatomic force constants shows that while anharmonicity suppresses thermal conductivity in pristine silicon, this effect weakens with vacancy concentration and reverses to enhance conductivity. These findings challenge conventional knowledge and provide new insights into thermal conduction in defective materials.

**Introduction**

The thermal conductivity of silicon containing defects, such as vacancies, interstitials, and voids, is of vital technological importance in many application areas including thermal management of electronics[1] and thermoelectrics[2,3]. For perfect silicon, its thermal transport behavior can be well described by phonon Boltzmann transport equation (BTE)[4,5], considering only phonon-phonon scattering mechanisms[6]. After defects are introduced, the reduction in thermal conductivity is conventionally attributed to the defect-phonon scattering[7-10]. A large amount of research has been focused on quantitatively assessing the defect scattering parameters[11-15]. The thermal conductivity of defective crystal is usually calculated by solving BTE considering phonon-defect scattering, with the assumption that vibrational frequencies and phonon velocities do not change with the defect concentration. However, crystal containing defects breaks the symmetry, which is an intermediate phase between crystal and glass. As a result, the particle-like propagation and scattering of vibrational modes may not fully capture heat transfer mechanism in defective crystals. Heat may also be transferred via wave-like tunnelling between vibrational modes. In addition, unit cells cannot be well defined in defective crystals, which may lead to the changes in vibrational frequencies and velocity operators. In this work, we explore thermal transport mechanisms beyond BTE and consider the vacancy induced changes in vibrational frequencies and velocity operators.

Recent advancements in thermal transport theories, such as the Wigner formulation of transport equation (WTE)[16,17] and the quasi-harmonic Green-Kubo (QHGK) approximation[18], have refreshed our understanding of heat conduction mechanisms in solids. These theories can consider the interplay between disorder, anharmonicity, and quantum Bose-Einstein statistics, naturally incorporating both particle-like and wave-like conduction mechanisms. Based on these theories, thermal conductivities of complex crystals[19-26] and glasses[27-29] have been successfully predicted. By considering both population and coherence contributions to the thermal conductivity, the transition from crystal-like to glass-like heat conduction has been observed through tuning atomic structures[21]. Although these new theories have made significant progress in understanding the thermal transport behaviors in crystals and glasses, their applications in defective crystals are rarely concerned. This work is devoted to applying WTE in the calculation of the thermal conductivity of silicon containing vacancies, seeking to gain a deeper understanding of the fundamental mechanisms that govern thermal conduction in defective crystals.

In addition to the consideration of coherence contribution to thermal conductivity, thermal transport theory can be further improved by incorporating anharmonicity renormalized frequencies, interatomic force constants (IFC), and higher-order scatterings between vibrational modes. Jain[19] found that anharmonicity renormalized IFC leads to more than a two-fold increase in the thermal conductivity of $Tl_3VSe_4$ at 300 K, which is offset by including 4-ph scattering processes. Xia *et al.*[20] showed that anharmonicity-induced hardening of frequencies reduces the scattering rates of acoustic modes, which compensates for the increase in scattering due to phonon population. These effects result in nearly temperature-independent thermal conductivities. Studies on silica by Zhu *et al.*[27] suggested that both anharmonic theory and anharmonic vibrational frequencies enhance thermal conductivity compared with harmonic theory and harmonic vibrational frequencies. These investigations indicate that the anharmonicity induced changes in vibrational properties should be considered as a factor that may influence the calculation results of thermal conductivities.

Conventionally, the impact of defects in crystal on heat conduction has been modeled as the introduction of additional scattering centers, which reduces thermal conductivity by shortening vibrational lifetimes. The vibrational frequencies and group velocities are assumed to be unaffected by the defects. However, group velocities are expected to play a significant role in thermal conduction, as indicated by the formulation of thermal conductivity[10]. Experimental work by Hanus *et al.*[30] showed that the speed of sound in PbTe decreases with the increase in the internal strain, which was found to be a dominate factor for the reductions in the thermal conductivity. Kargar *et al.*[31] observed a decrease in the velocity of acoustic phonons of $Al_2O_3$ crystals due to Nd atoms doping. These findings highlight the need for a reassessment of the assumption of constant group velocities in defective crystals, and a deeper investigation on how group velocities are influenced by defects.

In this work, we calculate the thermal conductivity of silicon with vacancies using WTE, which accounts for both particle-like propagation and wave-like tunneling of vibrational modes. By performing normal mode decomposition (NMD) on trajectories from molecular dynamics (MD) simulations, anharmonic vibrational frequencies and lifetimes are obtained, inherently incorporating all orders of anharmonicity. Through a renormalization method, anharmonic IFCs are also calculated. The effect of anharmonicity on the thermal conductivity is investigated considering both anharmonic frequencies and IFCs. Furthermore, we investigate how vacancies alter velocity operators and phonon lifetimes to elucidate the dominant mechanism behind vacancy-induced thermal conductivity reduction in silicon.

**Theory**

In this work, thermal conductivity is calculated by the formula derived from the Wigner transport equation[16]:

$$\begin{aligned} k^{\alpha\beta} = k_P^{\alpha\beta} + \frac{\hbar^2}{k_B T^2 V} \sum_{\mathbf{k}} \sum_{s,s'}^{s \neq s'} \frac{\omega(\mathbf{k})_s + \omega(\mathbf{k})_{s'}}{2} v^{\alpha}(\mathbf{k})_{s,s'} v^{\beta}(\mathbf{k})_{s',s} \\ \times \frac{\omega(\mathbf{k})_s f(\mathbf{k})_{0s}(f(\mathbf{k})_{0s}+1) + \omega(\mathbf{k})_{s'} f(\mathbf{k})_{0s'}(f(\mathbf{k})_{0s'}+1)}{4[\omega(\mathbf{k})_s - \omega(\mathbf{k})_{s'}]^2 + [\Gamma(\mathbf{k})_s + \Gamma(\mathbf{k})_{s'}]^2} \times [\Gamma(\mathbf{k})_s + \Gamma(\mathbf{k})_{s'}] \end{aligned} \quad (1)$$

where, $k_P^{\alpha\beta}$ is the populations term and the additional tensor is the coherences term. In the above equation, $\omega(\mathbf{k})_s$ is the frequency of $s$-mode, $v^{\alpha}(\mathbf{k})_{s,s'}$ is the velocity operator along $\alpha$ direction, $f(\mathbf{k})_{0s'}$ is the equilibrium Bose-Einstein distribution, $\Gamma(\mathbf{k})_s$ is the linewidth of $s$-mode that is the inverse of the lifetime $\tau(\mathbf{k})_s$, $\hbar$ is the reduced Planck constant, $k_B$ is the Boltzmann constant, $T$ is the temperature, and $V$ is the volume. $k_P^{\alpha\beta}$ can be computed by the solution of BTE. Under the relaxation time approximation, the calculation of $k_P^{\alpha\beta}$ can be simplified and the above equation becomes:

$$\begin{aligned} k^{\alpha\beta} = \frac{\hbar^2}{k_B T^2 V} \sum_{\mathbf{k}} \sum_{s,s'} \frac{\omega(\mathbf{k})_s + \omega(\mathbf{k})_{s'}}{2} v^{\alpha}(\mathbf{k})_{s,s'} v^{\beta}(\mathbf{k})_{s',s} \\ \times \frac{\omega(\mathbf{k})_s f(\mathbf{k})_{0s}(f(\mathbf{k})_{0s}+1) + \omega(\mathbf{k})_{s'} f(\mathbf{k})_{0s'}(f(\mathbf{k})_{0s'}+1)}{4[\omega(\mathbf{k})_s - \omega(\mathbf{k})_{s'}]^2 + [\Gamma(\mathbf{k})_s + \Gamma(\mathbf{k})_{s'}]^2} \times [\Gamma(\mathbf{k})_s + \Gamma(\mathbf{k})_{s'}] \end{aligned} \quad (2)$$

The velocity operator $\mathbf{v}(\mathbf{k})_{s,s'}$ is defined by:

$$v^{\beta}(\mathbf{k})_{s,s'} = e_{\kappa}^{\alpha*}(\mathbf{k})_s \nabla_{\mathbf{k}}^{\beta} \sqrt{D(\mathbf{k})}_{\kappa\alpha,\kappa'\alpha'} e_{\kappa'}^{\alpha'*}(\mathbf{k})_{s'}, \quad (3)$$

which can also be calculated by [16,32,33]:

$$\begin{aligned} \mathbf{v}(\mathbf{k})_{s,s'} = \frac{i}{\omega(\mathbf{k})_s + \omega(\mathbf{k})_{s'}} \sum_{\alpha,\beta} \sum_{l,\kappa,\kappa'} e_{\kappa}^{\alpha*}(\mathbf{k})_s \frac{\Phi_{\kappa'\kappa}^{\beta\alpha}(0,l)}{\sqrt{m_{\kappa} m_{\kappa'}}} \\ \times (\mathbf{R}_l + \mathbf{R}_{\kappa\kappa'}) e^{i\mathbf{k}\cdot(\mathbf{R}_l + \mathbf{R}_{\kappa\kappa'})} \\ \times e_{\kappa'}^{\beta}(\mathbf{k})_{s'} \end{aligned} \quad (4)$$

In the above equation, $\Phi_{\kappa'\kappa}^{\beta\alpha}(0,l)$ is the non-Hermitian force constants, $e_{\kappa}^{\alpha}(\mathbf{k})_s$ is the phonon eigenvector, $m_{\kappa}$ is the mass of atom $\kappa$, $\mathbf{R}_l$ is the position of cell $l$, and $\mathbf{R}_{\kappa\kappa'}$ is the distance between atom $\kappa$ and atom $\kappa'$ in a cell.

To obtain anharmonic vibrational frequencies and lifetimes, MD simulation-based NMD is carried out[34-36]. The velocities of each atom from MD simulations are first projected onto the normal mode:

$$Q(\mathbf{k},s,t) = \sum_{\alpha}^{3} \sum_{j}^{n} \sum_{l}^{N} \sqrt{\frac{m_j}{N}} u_{jl}^{\alpha}(t) e_j^{\alpha*}(\mathbf{k})_s \exp(-i\mathbf{k} \cdot \mathbf{r}_{lj}), \quad (5)$$

in which, $N$ is the total number of unit cells, $m_j$ is the mass of atom $j$, $u$ is the displacement, and $e_j^{\alpha}(\mathbf{k})_s$ is the phonon eigenvector. Then, Fourier transform of the time derivative of the normal mode is conducted to obtain the spectral energy density (SED):

$$\Psi(\mathbf{k},s,\omega) = \left| F(\dot{Q}(\mathbf{k},s,t)) \right|^2 = \left| \int_0^{+\infty} \dot{Q}(\mathbf{k},s,t) e^{-i\omega t} dt \right|^2. \quad (6)$$

The anharmonic vibrational frequencies and lifetimes are obtained through fitting the SED by a Lorentzian function:

$$\Psi(\mathbf{k},s,\omega) = \frac{C(\mathbf{k})_s}{\left[\omega - \omega^A(\mathbf{k})_s\right]^2 + \left[\Gamma(\mathbf{k})_s\right]^2}, \quad (7)$$

where, $\omega$ is the harmonic frequency, $\omega^A$ is the anharmonic frequency, and $\Gamma$ is the linewidth.

To consider the anharmonicity in the calculation of the thermal conductivity, both anharmonic frequencies and anharmonic interatomic force constants should be used. Anharmonic frequencies can be obtained through the above NMD method. Anharmonic force constants are calculated through a renormalization method described below. The relationship between force constants and dynamical matrix is:

$$\Phi_{0\kappa\alpha,l'\kappa'\alpha'} = \frac{\sqrt{m_\kappa m_{\kappa'}}}{N} \sum_{\mathbf{k}} D_{\kappa\alpha,\kappa'\alpha'}(\mathbf{k}) e^{-i\mathbf{k} \cdot (\mathbf{R}_{l'\kappa'}^0 - \mathbf{R}_{0\kappa}^0)}. \quad (8)$$

Considering that frequencies are eigenvalues of the dynamical matrix, the dynamical matrix can be computed by the following equation:

$$D_{\kappa\alpha,\kappa'\alpha'}(\mathbf{k}) = \sum_s e_\kappa^{\alpha}(\mathbf{k})_s \omega_{\mathbf{k}s}^2 e_{\kappa'}^{\alpha'*}(\mathbf{k})_s. \quad (9)$$

The anharmonic force constants are obtained through Eq. (8), in which the dynamical matrix is calculated by substituting anharmonic frequencies into Eq. (9).

**Results**

**Thermal conductivity**

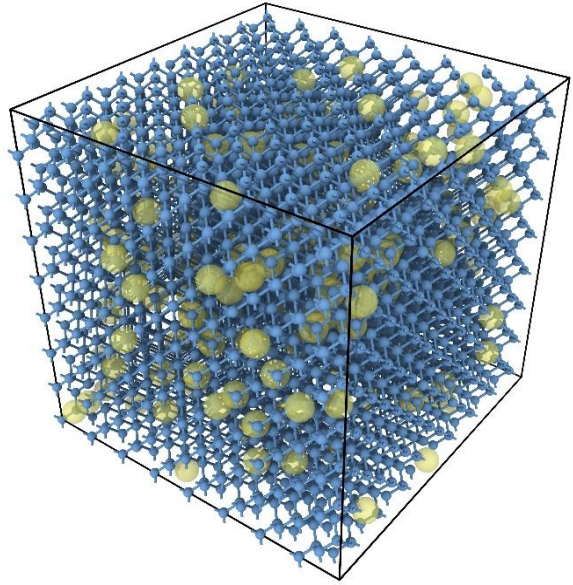

**Fig. 1 Schematic of vacancy defects in silicon.** Silicon atoms are depicted as blue spheres, while vacancies — randomly distributed within the silicon crystal lattice — are represented as semi-transparent yellow spheres.

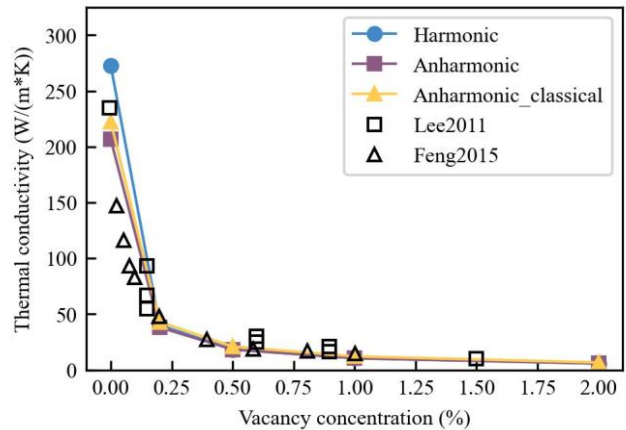

**Fig. 2 Thermal conductivity of silicon as a function of vacancy concentration.** The filled points represent data calculated by this work using Eq. (1), while the hollow points correspond to data from MD simulations in the literatures[7,9].

Vacancy defects are randomly distributed in silicon lattice as shown in Fig. 1. The variation of the thermal conductivity with respect to the vacancy concentration is depicted in Fig. 2. The increase in vacancy concentration leads to a rapid decrease in the thermal conductivity. Even with 0.2% vacancies, the thermal conductivity of the silicon decreases by 85.25%. The presence of 2.0% vacancies results in a 97.88% decrease in the thermal conductivity compared to perfect silicon. The results of this work are in good agreement with the results of non-equilibrium MD simulations by

Lee *et al.*[7] and Green-Kubo methods by Feng *et al.*[9] using Tersoff potential. Although more accurate machine-learning-based interatomic potentials have been developed for predicting the thermal conductivity in pure silicon and silicon with vacancy defects[37], the primary focus of this work is to investigate the mechanisms by which vacancy defects impede thermal transport in silicon, rather than to obtain accurate thermal conductivity values. Therefore, we used the classical Tersoff potential to ensure computational efficiency and facilitate benchmarking against existing studies. To exclude the influence of artificial periodicity, thermal conductivities of both pure silicon and defective silicon with different model sizes are calculated using the Green-Kubo method based on MD simulations as shown in the supporting information. It is found that 5×5×5 supercell is sufficient to achieve converged thermal conductivity.

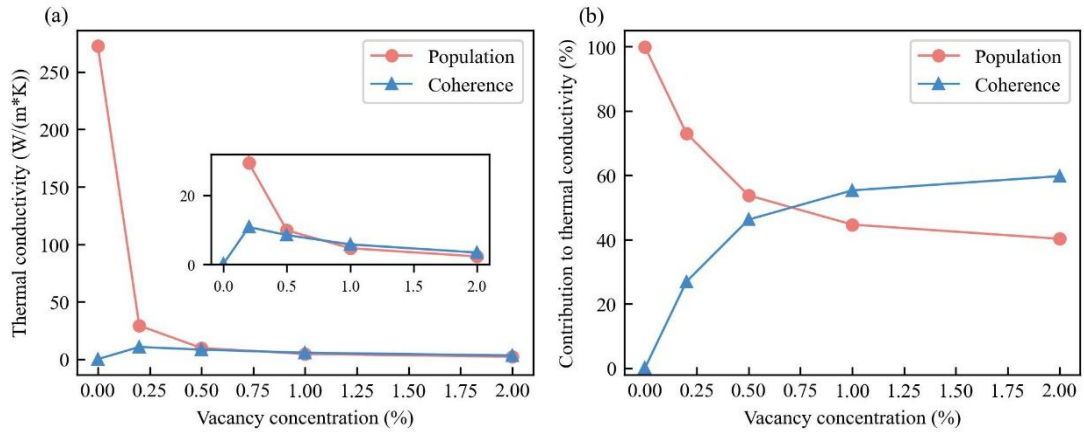

**Fig. 3 Thermal conductivity decomposed into the populations and coherences terms.** (a) Variation of the thermal conductivity contributed by the populations term and coherences term with respect to the vacancy concentration, respectively. (b) Relative contribution of the populations term and coherences term to the total thermal conductivity.

Total thermal conductivity is decomposed into the contributions of populations and coherences. Since the crystal is treated as a large unit cell and only Gamma point is considered in this work, exactly degenerate states do not exist. Quasi-degenerate states with frequency differences smaller than 0.1 THz are regarded as degenerate states that contribute to the populations term. As shown in Fig. 3(a), the thermal conductivity contributed by populations term decreases with the increase in the vacancy concentration, while the contribution from the coherences term first increases then decreases. Figure 3(b) shows that populations term dominates when the vacancy concentration is below 1.0%. In perfect crystal, populations term contributes 99.91% of the total thermal conductivity. However, when the vacancy concentration exceeds 1.0%, coherences term dominates. For silicon with 2.0% vacancy concentration, populations term contributes only 34.18% of the total thermal conductivity, while the coherences term contributes 65.82%. This indicates that, with the increase in vacancy concentration, heat transfer transits from particle-like propagation of phonon wavepacket to wave-like tunnelling.

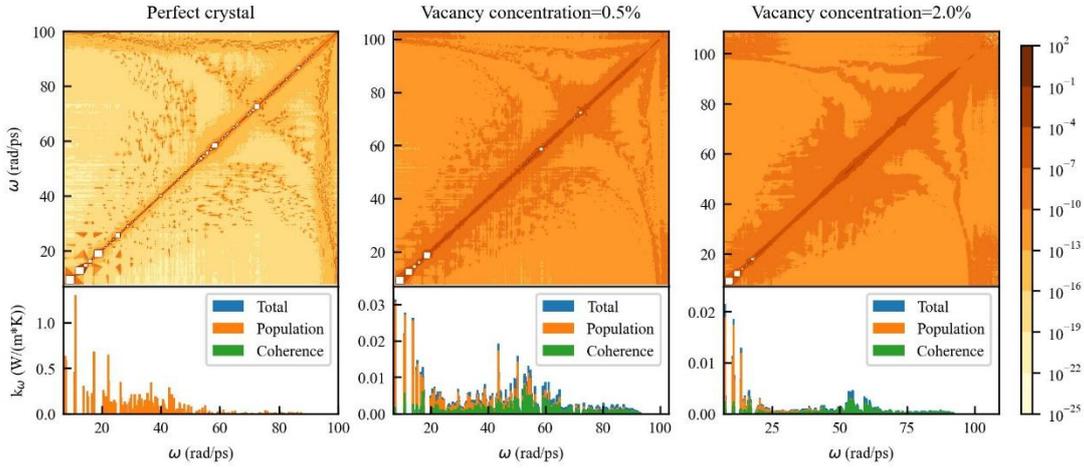

**Fig. 4 Two-dimensional and one-dimensional frequency-dependent thermal conductivities.**

Fig. 4 shows the frequency-dependent thermal conductivities, $k_{\omega\omega'}$ and $k_\omega$, for perfect silicon and silicon containing vacancies, respectively. For perfect crystal, $k_{\omega\omega'}$ is mainly distributed near the diagonal and drops sharply away from the diagonal. With the increase in $\omega$, populations term of $k_\omega$ decreases rapidly. These results indicate that the thermal conductivity of perfect silicon is primarily contributed by quasi-degenerate states at low frequencies. For silicon containing vacancies, the difference between $k_{\omega\omega'}$ near the diagonal and away from the diagonal becomes smaller compared with that in perfect silicon. Coherences term in vacancy defected silicon is non-negligible, that mainly originates from the medium frequency spectrum of vibrational states, as indicated by the green bars in Fig. 4.

Fig. 5 shows the contribution of coherences between vibrational modes with frequency difference smaller than a specified frequency cutoff to the total thermal conductivity. For perfect crystal, the curve nearly saturates at a frequency cutoff of 0.1 THz. This indicates the dominant role of coupling between quasi-degenerate states in the thermal conduction of perfect silicon. As the vacancy concentration increases, the frequency cutoff required to achieve saturation also rises. This suggests that coherences between vibrational modes with larger frequency differences can participate in the thermal conduction as the vacancy concentration increases. However, the contribution of these coherences diminishes with increasing frequency differences, as evidenced by the decreasing slope of the curve with the frequency cutoff. Moreover, Fig. 5 reveals that even in silicon with 2.0% vacancies, 99% of the thermal conductivity originates from coherences between vibrational modes with frequency differences no larger than 1.3 THz. This indicates that in silicon containing vacancies, the coherences term of thermal conductivities is mainly contributed by the coherences between vibrational modes with small frequency differences. This finding is in agreement with previous investigations that coherences contribution arising from structural disorder is predominantly due to couplings between states with small frequency differences [16,20].

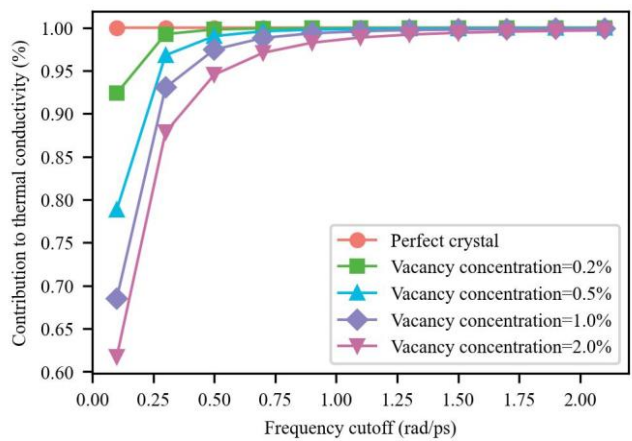

**Fig. 5 Sub-cutoff coherence contribution to thermal conductivity.** The coherence contributions between vibrational modes with frequency differences smaller than the frequency cutoff are divided by total thermal conductivities.

**Anharmonicity**

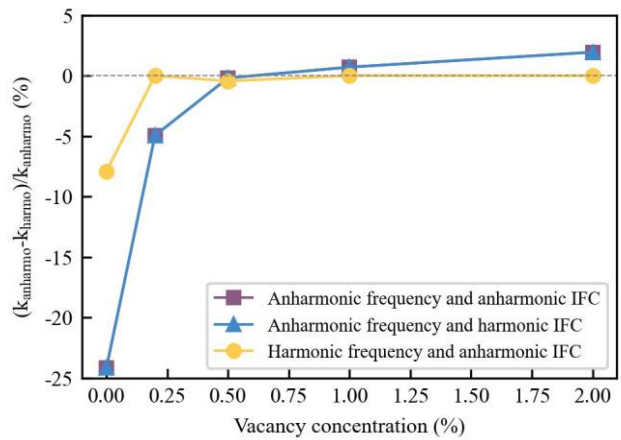

**Fig. 6 The relative differences between thermal conductivities arising from anharmonic and harmonic treatments of vibrational frequencies and interatomic force constants (IFCs).**

Thermal conductivities calculated using anharmonic frequencies and IFCs are compared with those obtained through harmonic approximations. The relative difference between the results of these two methods is shown by the purple line in Fig. 6. For perfect silicon, anharmonic effects induce a 24.18% suppression of thermal conductivity. Notably, the influence of anharmonicity progressively attenuates with increasing vacancy concentration below 1.0%. Intriguingly, when the vacancy concentration reaches 1.0%, anharmonic contributions enhance thermal conductivity. This effect exhibits progressive amplification with increasing vacancy concentration, which is accompanied by the increase in the structural disorder. This phenomenon aligns with our prior observations in fully disordered amorphous silica systems[27], wherein anharmonic interactions elevate thermal transport capabilities. These collective findings demonstrate that the functional role of anharmonicity in heat conduction undergoes a reversal - transitioning from detrimental suppression to beneficial enhancement - as material disorder evolves beyond critical threshold.

A decoupling analysis is performed to isolate the individual effects of anharmonic frequency renormalization and anharmonic IFC on thermal conductivities. The investigation proceeds through two controlled computational experiments. First, while maintaining harmonic IFCs, thermal conductivity values computed with anharmonic frequencies are compared with those calculated using harmonic frequencies, as shown by the blue line in Fig. 6. Second, thermal conductivity values obtained using anharmonic IFCs under fixed harmonic frequencies are compared with those derived from harmonic IFCs, as shown by the yellow line in Fig. 6. The blue line almost overlaps with the purple line. This reveals that the thermal conductivity modification induced solely by anharmonic vibrational frequencies exhibits comparable magnitude to the combined effects of anharmonic frequencies and anharmonic IFCs. Comparison between the yellow line and blue line demonstrates that implementing only anharmonic IFCs produces a significantly weaker thermal conductivity variation than isolated anharmonic frequency adjustments. These findings demonstrate that frequency renormalization through anharmonic effects constitutes the primary mechanism governing anharmonicity-driven thermal conductivity modifications.

**Mechanism**

To uncover the mechanism for heat transfer in silicon containing vacancies, Eq. (1) can be rewritten into the following form:

$$k^{\alpha\beta} = \frac{1}{V}\sum_{\mathbf{k}}\sum_{s,s'} C(\mathbf{k})_{s,s'} V^{\alpha}(\mathbf{k})_{s,s'} V^{\beta}(\mathbf{k})_{s,s'} \tau(\mathbf{k})_{s,s'}, \tag{10}$$

where,

$$C(\mathbf{k})_{s,s'} = \frac{\omega(\mathbf{k})_s + \omega(\mathbf{k})_{s'}}{4}\left[\frac{C(\mathbf{k})_s}{\omega(k)_s} + \frac{C(\mathbf{k})_{s'}}{\omega(k)_{s'}}\right] \quad (11)$$

and

$$\tau(\mathbf{k})_{s,s'} = \frac{\left[\Gamma(\mathbf{k})_s + \Gamma(\mathbf{k})_{s'}\right]/2}{\left[\omega(\mathbf{k})_s - \omega(\mathbf{k})_{s'}\right]^2 + \left[\Gamma(\mathbf{k})_s + \Gamma(\mathbf{k})_{s'}\right]^2/4} \quad (12)$$

are defined as the two-mode heat capacity and lifetime [38], respectively. It has been verified that the effect of the vacancy concentration on the two-mode heat capacity is negligible. Therefore, it is sufficient to explore the vacancy-induced changes in the velocity operators and two-mode lifetimes to uncover the mechanisms of thermal transport degeneration in silicon containing vacancies.

Two-dimensional frequency dependent velocity operators that have been diagonalized in degenerate subspaces are shown in Fig. 7. The contour map of velocity operators is similar to the contour map of thermal conductivities shown in Fig. 4. The maximum velocity is primarily concentrated around the diagonal. Away from the diagonal, only localized regions have relatively high velocities. With the increase in the vacancy concentration, velocities around the diagonal decrease, while those away from the diagonal increase.

To quantitatively demonstrate how velocities vary with vacancy concentration, the velocities along the diagonal, as well as those deviating from the diagonal by 0.5 rad/ps and 1 rad/ps, are plotted in Fig. 8. The velocity along the diagonal (offset = 0) that contributes to the populations term decreases rapidly as vacancy concentration increases. In contrast, the velocity deviating from the diagonal by 0.5 rad/ps first increases and then decreases with increasing vacancy concentration. The velocity deviating from the diagonal by 1.0 rad/ps increases continuously with vacancy concentration. For perfect silicon, the average velocity along the diagonal is 21.46 Å/THz, which decreases to 0.02 Å/THz for a deviation of 0.5 rad/ps. However, for silicon with 2.0% vacancies, the average velocity only decreases from 4.32 Å/THz to 1.07 Å/THz when the frequency deviates from the diagonal by 0.5 rad/ps. These results indicate that the diagonal elements dominate the velocity operator, and this dominance decreases as vacancy concentration increases. The difference between the diagonal and off-diagonal elements of the velocity operator shrinks with the increase in the vacancy concentration. These findings reveal that vacancy-induced changes in the velocity operators may serve as a key contributor to thermal conductivity reductions, and part of the heat conduction transits from particle-like propagation to wave-like tunnelling as vacancy concentration increases.

Fig. 8 also shows how velocity varies with frequency. The diagonal velocity decreases as the frequency increases. For the velocity deviating from the diagonal by 0.5 rad/ps and 1 rad/ps, it first increases to values oscillating around a constant at frequency of approximate 20 rad/ps, and then decreases at frequency of around 85 rad/ps. These results indicate that the populations term, associated with diagonal velocities, is mainly contributed by the low-frequency modes, while the coherences term, related to off-diagonal velocities, is mainly contributed by modes with intermediate frequencies.

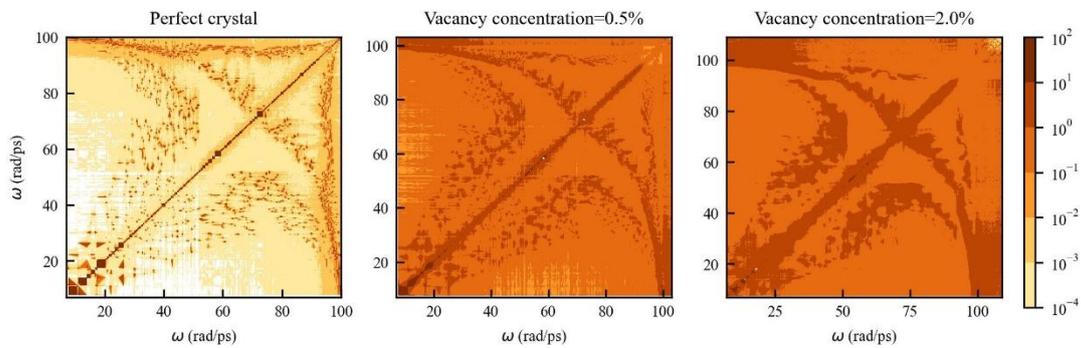

**Fig. 7 Two-dimensional frequency dependent velocity operators that have been diagonalized in the degenerate subspaces.**

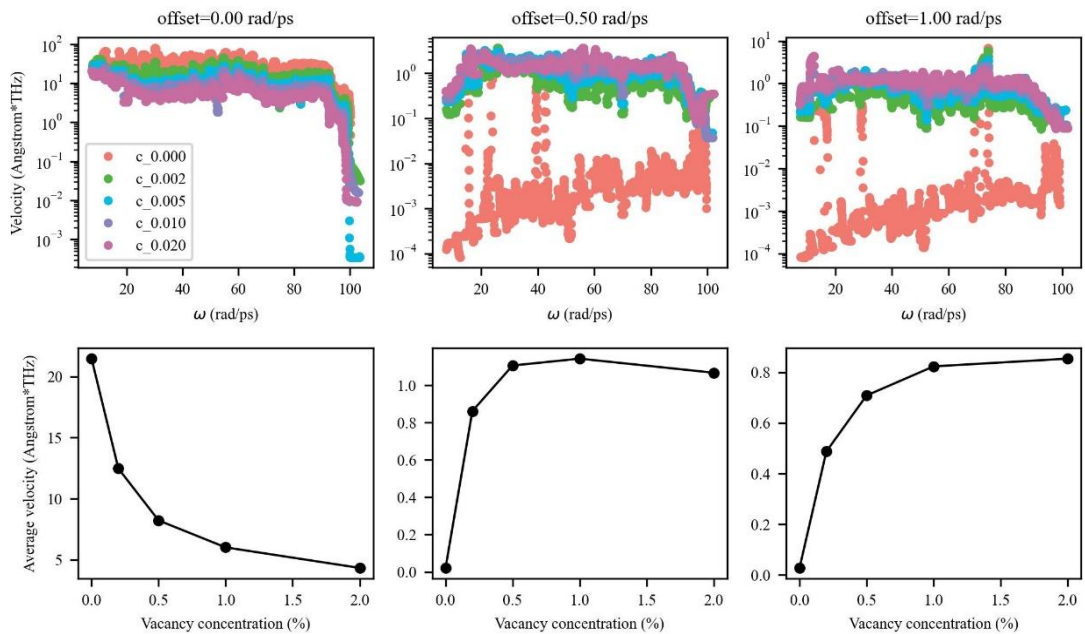

**Fig. 8 Velocity operators along the diagonal (offset = 0), and those deviating from the diagonal by 0.5 rad/ps and 1 rad/ps.** The upper figures show the variation of the velocity with frequency,

while the lower figures show the variation of the frequency-averaged velocity with vacancy concentration. The frequency-dependent velocity is calculated by the formula $v_\omega = \sum_s v_s \delta(\omega_s - \omega) / \sum_s \delta(\omega_s - \omega)$ with the dela function approximated by Gaussian function.

The contour map of the two-mode lifetimes calculated by Eq. (12) is plotted in Fig. 9. The maximum two-mode lifetime is concentrated along the diagonal. As the distance from the diagonal increases, the two-mode lifetime decreases. However, it is difficult to discern from the contour map how the two-mode lifetimes vary with the vacancy concentration and frequency.

To quantitatively demonstrate how two-mode lifetimes vary with the vacancy concentration and frequency, the two-mode lifetimes along the diagonal, as well as those deviating from the diagonal by 0.5 rad/ps and 1 rad/ps, are plotted in Fig. 10. Both the diagonal and off-diagonal elements of the two-mode lifetimes decrease as the vacancy concentration increases. With the increase in the frequency, the diagonal lifetimes decrease. For off-diagonal lifetimes, they first increase to values around a constant, and then decrease with the frequency. This behavior is similar to the variation of velocity with the frequency described above, which further supports the finding that low-frequency modes dominate the particle-like propagation and the coherences term of wave-like tunnelling is mainly contributed by modes with intermediate frequencies.

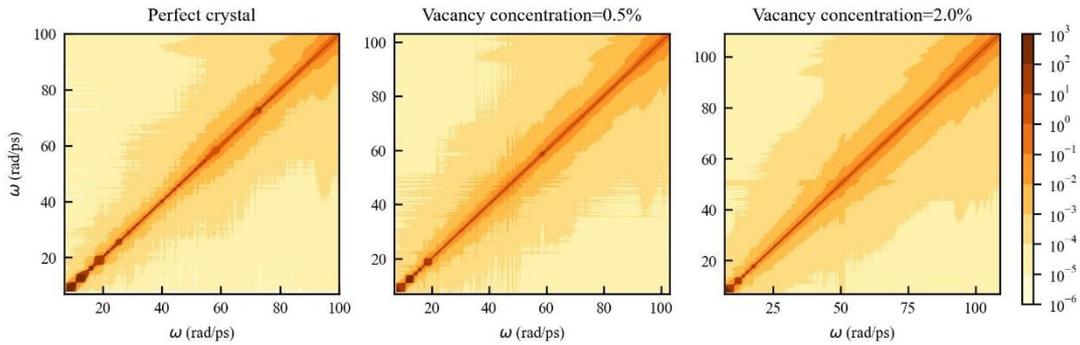

**Fig. 9 Two-dimensional frequency dependent two-mode lifetimes.**

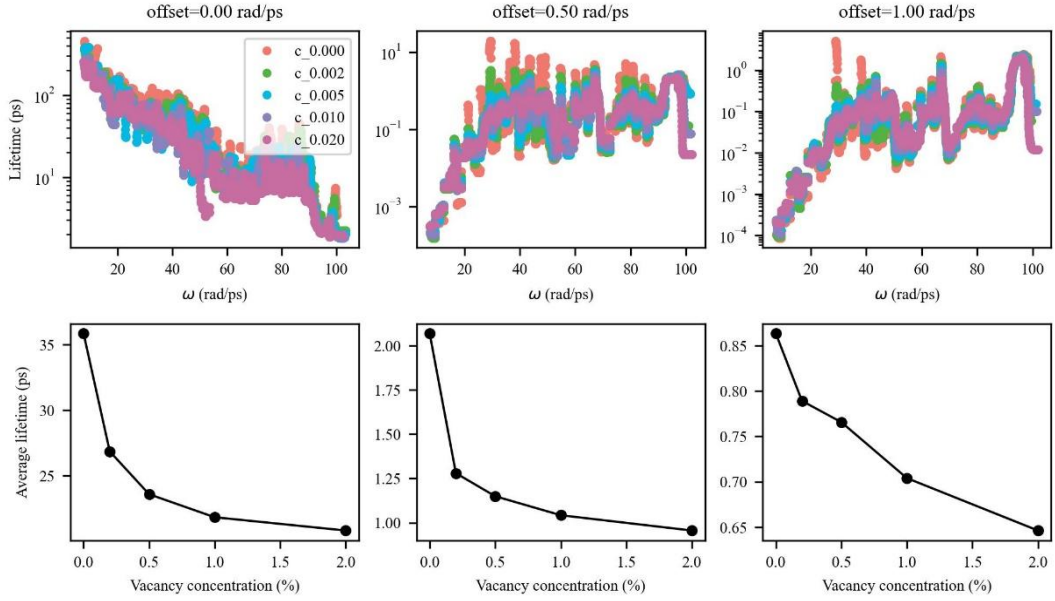

**Fig. 10 Two-mode lifetimes along the diagonal (offset = 0), and those deviating from the diagonal by 0.5 rad/ps and 1 rad/ps.** The upper figures show the variation of the two-mode lifetime with frequency, while the lower figures show the variation of the frequency-averaged two-mode lifetime with vacancy concentration. The frequency-dependent lifetime is calculated by the formula $\tau_\omega = \sum_s \tau_s \delta(\omega_s - \omega) / \sum_s \delta(\omega_s - \omega)$ with the dela function approximated by Gaussian function.

The above computational results demonstrate vacancy concentration-dependent behaviors in both velocity operators and phonon lifetimes, suggesting dual mechanisms for thermal conductivity reduction. To decouple these effects, we perform comparative calculations of thermal conductivities as a function of vacancy concentration: (1) varying velocity operators under fixed lifetimes, and (2) changing lifetimes with constant velocity operators. The results are shown in Fig. 11. The vacancy-induced changes in velocity operators exhibit larger impacts on thermal conductivity reductions compared to lifetime suppression mechanisms. This finding challenges the long-standing defect scattering paradigm in phonon transport theory.

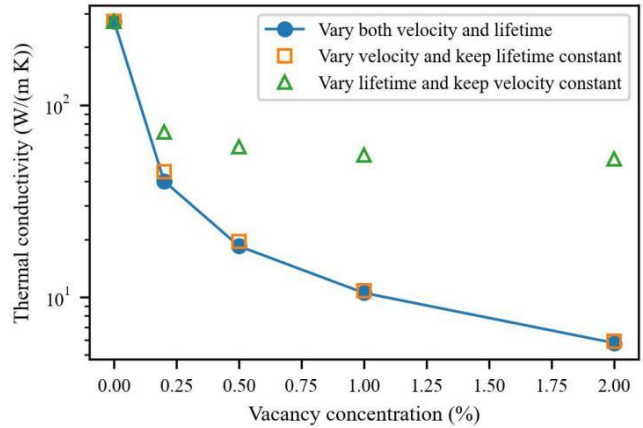

**Fig. 11 Comparison of vacancy effects on thermal conductivities via independent changes of velocity operators and vibrational lifetimes.** The hollow orange points show thermal conductivity vs. vacancy concentration calculated by varying velocity operators under fixed lifetimes, while the hollow greens points obtained by changing lifetimes with constant velocity operators.

**Conclusions**

In conclusion, we investigated the thermal conduction in vacancy defected silicon, considering both the particle-like propagation of vibrational modes and wave-like tunnelling between modes. Our analyses incorporate anharmonic vibrational frequencies, interatomic force constants, and vacancy-induced changes in both velocity operators and lifetimes. It is found that although the particle-like propagation dominates in defect-free silicon, wave-like tunnelling becomes non-negligible in silicon containing vacancies. As vacancy concentration increases, wave-like tunnelling becomes increasingly important. Further investigations on the frequency-dependent thermal conductivities, velocity operators, and lifetimes reveal that particle-like propagation is primarily contributed by the low-frequency modes, while the coherences term of wave-like tunnelling is mainly contributed by modes with intermediate frequencies. Considering anharmonic frequencies and IFCs in the thermal conductivity calculations, we found that anharmonic effects suppress thermal conductivity in pristine silicon. However, increasing vacancy concentration progressively attenuates this anharmonic suppression, with a critical transition occurring at 1% concentration where the effect shifts from conductivity reduction to enhancement. To elucidate thermal transport mechanisms in vacancy-defected silicon, we systematically investigate vacancy-induced modifications in both velocity operators and vibrational lifetimes. Our analyses demonstrate that the observed thermal conductivity reduction arises predominantly from altered velocity operators, rather than lifetime shortening. This finding challenges the conventional knowledge that defects reduce thermal conductivity primarily through enhanced phonon scattering, suggesting that current defect engineering strategies for thermal management may require reassessment. The revealed dominance of phonon velocity reduction mechanism provides new insights for developing thermal design methodologies.

**Methods**

MD simulations are performed using LAMMPS package[39] to calculate the vibrational frequencies and lifetimes through lattice dynamics calculations and NMD method. The interaction energy between silicon atoms is calculated by Tersoff interatomic potential[40]. The domain size is set to 8×8×8 conventional unit cell, with periodic boundary conditions applied in all three directions. Vacancies of concentrations 0.2%, 0.5%, 1.0% and 2.0% are introduced by randomly deleting silicon atoms.

All the simulations are performed with periodic boundary conditions and a timestep of 0.5 fs at 300 K. The system is first equilibrated for 1.5 ns in NPT ensemble with the last 0.5 ns for averaging the box size. After setting the box size to the averaged value, the system is then simulated for 0.5 ns in NVT ensemble and 1 ns in NVE ensemble. The last frame of the above simulations is processed through an energy minimization for obtaining the equilibrium configuration.

Based on the equilibrium configuration, the system is equilibrated for 0.5 ns in NVT ensemble and 0.5 ns in NVE ensemble. Then, the velocities of each atom are dumped in NVE ensemble for a time span of 5 ns, which are used for performing NMD.

Harmonic force constants are also calculated based on the equilibrium configuration. With the harmonic force constants and treating the total simulation box as a unitcell, lattice dynamics calculations are performed only at the gamma point using the Phonopy code[41] for obtaining the harmonic vibrational frequencies and eigenvectors.

**Acknowledgement**

This work was supported by National Natural Science Foundation of China (Grant No. 12005019 and Grant No. 52306102).